\pdfoutput=1
\documentclass[aps,prl,reprint,superscriptaddress]{revtex4-1}
\usepackage{graphicx,color}
\usepackage{amsmath}
\usepackage{inputenc}
\usepackage[T1]{fontenc}
\usepackage{ulem}
\usepackage{lmodern}
\usepackage{amsfonts}

\newcommand{\vect}[1]{\mathbf{#1}}
\newcommand{\uv}[1]{\hat{\mathbf{#1}}}
\newcommand{\degsym}{^{\circ}}
\newcommand{\mh}{\mathfrak{h}}
\newcommand{\mk}{\mathfrak{K}}
\newcommand{\ml}{\mathfrak{l}}
\begin{document}

%

\title{Simulating complex crystal structures using the phase-field crystal model}

\author{Eli Alster}
\affiliation{Department of Chemical and Biological Engineering, Northwestern University, Evanston, Illinois 60208, USA}
\affiliation{Department of Materials Science and Engineering, Northwestern University, Evanston, Illinois 60208, USA}
\author{David Montiel}
\affiliation{Department of Materials Science and Engineering, University of Michigan, Ann Arbor, MI 48109, USA}
\author{Katsuyo Thornton}
\affiliation{Department of Materials Science and Engineering, University of Michigan, Ann Arbor, MI 48109, USA}
\author{Peter W. Voorhees}
\email[]{p-voorhees@northwestern.edu}
\affiliation{Department of Materials Science and Engineering, Northwestern University, Evanston, Illinois 60208, USA}

\date{\today}

\begin{abstract}
We introduce a phase-field crystal model that creates an array of complex three- and two-dimensional crystal structures via a numerically tractable three-point correlation function. The three-point correlation function is designed in order to energetically favor the principal interplanar angles of a target crystal structure. This is achieved via an analysis performed by examining the crystal's structure factor. This approach successfully yields energetically stable simple cubic, diamond cubic, simple hexagonal, graphene layers, and CaF$_2$ crystals. To illustrate the ability of the method to yield a particularly complex and technologically important crystal structure, we show how this three-point correlation function method can be used to generate perovskite crystals.
\end{abstract}

\maketitle
Multiscale phenomena in materials are notoriously difficult to model because they elude conventional techniques like molecular dynamics, continuum mechanics, and phase-field theory. This is a problem because most interesting phenomena span multiple orders of magnitude in both time and length scales. One promising method for multiscale simulation of crystalline materials is the phase-field crystal (PFC) method. The PFC method utilizes a free energy functional that is averaged over the time scale of atomic vibrations but retains patterns associated with lattice structures as equilibrium states \cite{Elder2007}. It reproduces Read-Shockley grain boundary energies \cite{Elder2002} and has been used to examine graphene grain boundary structure \cite{Hirvonen2016a}, to model step-flow growth from a supersaturated vapor \cite{Schwalbach2013}, and to study electromigration in metal interconnects \cite{Wang2016}, among other applications \cite{Emmerich2012}.

A major shortcoming of the PFC method is the limited number of crystal structures it can simulate. Progress in the PFC community has largely been made by considering various phenomenological forms for the Fourier transform of the two-point correlation function, $\hat{C}_2$ \cite{Greenwood2010, Greenwood2011}. For a $\hat{C}_2$ containing up to three peaks, there exist two-dimensional (2D) PFC models for all five Bravais lattices \cite{Mkhonta2013} and various chiral phases \cite{Mkhonta2016}. In three-dimensions, PFC models with this $\hat{C}_2$ can form simple cubic \cite{Greenwood2011}, face-centered cubic (fcc) \cite{Greenwood2011}, and diamond cubic structures \cite{Chan2015}. Consequently, applications of three-dimensional (3D) PFC models have been dominated by simulations employing body-centered cubic (bcc)  \cite{Wu2007, Jaatinen2009, Berry2014, Wu2016, Alster2017} and fcc \cite{Wu2010, Berry2012, Fallah2013b, Berry2014, Fallah2016} crystal structures. Obviously, there exist many more crystal structures, and no attempts have been made to describe a crystal as complex as perovskite, a deficit that we address in this Letter. 

The challenge of producing complex crystal structures in PFC models is similar to that faced by the self-assembly community. Their goal is to solve the so-called ``inverse'' statistical mechanics problem: how to design interaction potentials between discrete particles such that a given structure is a global energy minimum. They have also found this task to be non-trivial \cite{Rechtsman2006, Rechtsman2007, Edlund2011, Barkan2014}.

Traditionally, the single-component PFC free energy functional is expressed as a combination of one-body and two-body interactions. Namely,
\begin{align} \label{eq:e2}
F[n] = F_{1}[n] + F_{2}[n] = &\int_V \left [ \frac{1}{2}n^2 - \frac{1}{6}n^3 + \frac{1}{12}n^4 \right ]d\vect{r} \nonumber \\
- \frac{1}{2} &\int_V n(\vect{r}) C_2 \ast n \text{ } d\vect{r} \text{,}
\end{align}
where $F$ is a nondimensionalized free energy, $n$ is a nondimensionalized density, $F_{1}$ is the ideal free energy term, $F_2$ is the energy from two-point interactions, $V$ is the system volume, $C_2$ is an isotropic two-point correlation function, and $C_2 \ast n \equiv \int_{V'} C_2(|\vect{r} - \vect{r}'|) n(\vect{r}') d\vect{r}'$ \cite{Elder2007, Greenwood2010, Greenwood2011}. Since this free energy functional is rotationally and translationally invariant, it is not a trivial task to design a $C_2$ function that produces the desired crystal structure as an energy minimum. In fact, symmetry considerations suggest that most forms of $C_2$ will only result in a bcc or lower-dimensional structure \cite{Alexander1978}.

In an effort to derive an improved model for graphene, Seymour \textit{et al.} added the energy due to three-point correlations \cite{Seymour2016},
\begin{equation} \label{eq:esum}
F[n] = F_1 + F_2 + F_3 \text{,}
\end{equation}
where $F_1$ and $F_2$ are the same as in Eq. \ref{eq:e2} and
\begin{equation} \label{eq:e3}
F_3[n] = - \frac{1}{6} \iiint n(\vect{r}) C_3(\vect{r} - \vect{r}', \vect{r} - \vect{r}'') n(\vect{r}')n(\vect{r}'') d\vect{r}d\vect{r}'d\vect{r}'' \text{.}
\end{equation}
Although in general the calculation of $F_3$ is of computational complexity $O(N^3)$, if the $C_3$ function is of the form
\begin{equation} \label{eq:C3sep}
C_3(\vect{r}_1, \vect{r}_2) = \sum_i C_3^{(i)}(\vect{r}_1)C_3^{(i)}(\vect{r}_2) \text{,}
\end{equation}
where $\vect{r}_1 \equiv \vect{r} - \vect{r}'$ and $\vect{r}_2 \equiv \vect{r} - \vect{r}''$, the free energy and evolution equations simplify into a number of convolutions of $O(N \log N)$ computational complexity that are easily computed via the fast-Fourier transform. Seymour \textit{et al.} proposed a single length scale, 2D real space form for $C_3^{(i)}$ that yields equilibrium states with a specified bond angle when the bond angle, $\theta$, satisfies the relation
\begin{equation} \label{eq:restrictAngle}
360\degsym \textit{ mod } \theta = 0 \text{,}
\end{equation}
for example $60 \degsym$, $90 \degsym$, and $120 \degsym$ \cite{Seymour2016}. Although this yielded an improved 2D graphene model \cite{Hirvonen2016a}, the model was not flexible enough to generate any new crystal structures in either two or three dimensions  \cite{Seymour2016, SeymourPersonal}. 

In this Letter, we introduce a form for $C_3$ that stabilizes angles between specified crystallographic planes and can include multiple length scales, multiple preferred angles, and angles not restricted by Eq. \ref{eq:restrictAngle}. Then, we will discuss how to choose parameters in order to yield energy-minimizing single-component crystal structures. We show that this method can produce a wide array of energy-minimizing crystal structures, from simple cubic (e.g., Po \cite{Beamer1946}) and diamond cubic (e.g., C-diamond, Si, $\alpha$-tin \cite{Mermin2014}), to graphene layers and disordered CaF$_2$ (e.g., the structure of the $\theta'_c$ phase of Al$_2$Cu, which is commercially very important for strengthening in aluminum alloys \cite{Wolverton2001}). Additionally, it produces an unnamed crystal structure corresponding to the $X$ atoms in $ABX_3$ perovskite (a structure we will call $X_3$), which is necessary for modeling perovskite.

Finally, as a capstone demonstration of the method, we combine the $X_3$ and simple cubic models to generate a perovskite crystal structure. Such compounds are of great interest due to applications ranging from high efficiency solar cells \cite{Yin2015, Abate2016} to light-emitting diodes \cite{Tan2014}. However, the development of perovskite microstructure is governed by phenomena that occur on a diffusional time scale, not the nanoseconds afforded by molecular dynamics. Consequently, this model will provide a new avenue to investigate multiscale phenomena in these important materials. 

The ansatz we use for $C_3$ is
\begin{equation} \label{eq:C3Basic}
\hat{C}_3(\vect{k}_1, \vect{k}_2) = \beta^2 R(k_1) R(k_2)\sum_{l=0}^{l_{\text{max}}} \alpha_l P_l(\uv{k}_1 \cdot \uv{k}_2) \text{,}
\end{equation}
where $\hat{C}_3$ is the Fourier transform of the three-point correlation function, $k_i = |\vect{k}_i|$, $\uv{k}_i = \vect{k}_i/|\vect{k}_i|$, $\beta$ is an interaction strength parameter, $R(k)$ is a real radial function, $P_l$ are the Legendre polynomials, and $\alpha_l$ are constant coefficients. As is explained later, $\alpha_l$ should be interpreted as determining the preferred interplanar angles. Because $\hat{C}_3(\mathcal{R}\vect{k}_1, \mathcal{R}\vect{k}_2) = \hat{C}_3(\vect{k}_1, \vect{k}_2)$ for any rotation matrix $\mathcal{R}$, $\hat{C}_3$ is rotationally invariant and so is $C_3$. By keeping the free energy rotationally invariant, it is possible to study phenomena such as solid-liquid interfaces and grain boundary energies as a function of misorientation, which would be impossible otherwise.

This ansatz was chosen not only because it is rotationally invariant but also because the Legendre polynomials are both separable and form a complete orthogonal set. More explicitly, the separability of the Legendre polynomials means that
\begin{equation}
P_l(\uv{k}_1 \cdot \uv{k}_2) = \sum_{m=-l}^{l} \frac{4 \pi}{2l+1} Y_{lm}(\uv{k}_1)Y_{lm}(\uv{k}_2) \text{,}
\end{equation}
where $Y_{lm}$ are the normalized real spherical harmonics \cite{harmonics}. Thus, $\hat{C}_3$ can be written as a sum of products of two-point correlation functions, i.e.,
\begin{equation} \label{eq:C3FS}
\hat{C}_3(\vect{k}_1, \vect{k}_2) = \sum_{l=0}^{l_{\text{max}}} \alpha_l (-1)^l \sum_{m=-l}^{l} \hat{C}^{(lm)}(k_1, \uv{k}_1) \hat{C}^{(lm)}(k_2, \uv{k}_2) \text{,}
\end{equation}
where
\begin{equation} \label{eq:ClmFS}
\hat{C}^{(lm)}(k,\uv{k}) \equiv (-i)^l \sqrt{\frac{4 \pi}{2l+1}} \beta R(k)Y_{lm}(\uv{k}) \text{.}
\end{equation}
The factor of $(-1)^l$ in Eq. \ref{eq:C3FS} was introduced to cancel the phase factors, $(-i)^l$, in Eq. \ref{eq:ClmFS} so that $C^{(lm)}$ is real (see Supplementary Materials for details). 

Substituting the inverse Fourier transform of Eq. \ref{eq:C3FS} into Eq. \ref{eq:e3} results in
\begin{equation}
F_3 = -\frac{1}{6}\sum_{l=0}^{l_{\text{max}}} \alpha_l (-1)^l \sum_{m=-l}^{l} \int n(\vect{r}) \bigg (C^{(lm)} \ast n \bigg )^2 d\vect{r}
\end{equation}
and
\begin{align}
\frac{\delta F_3}{\delta n} = -&\sum_{l=0}^{l_{\text{max}}} \frac{\alpha_l (-1)^l}{6} \sum_{m=-l}^{l} \bigg \{(C^{(lm)} \ast n)^2 \nonumber \\
& + 2(-1)^l C^{(lm)} \ast [n(C^{(lm)} \ast n)] \bigg \} \text{,}
\end{align}
since $C^{(lm)}(-\vect{r}) = (-1)^l C^{(lm)}(\vect{r})$ by the parity property of real spherical harmonics (note the similarity between these expressions and Eq. 10 and Eq. 42 respectively from \cite{Seymour2016}).

Since the Legendre polynomials form a complete orthogonal set, if 
\begin{equation} \label{eq:Bx}
B(x) \equiv \sum_{l=0}^{l_{\text{max}}} \alpha_l P_l(x) \text{,}
\end{equation}
then each $\alpha_l$ is given by
\begin{equation} \label{eq:alphalGen}
\alpha_l = \frac{2l+1}{2} \int_{-1}^{1} B(x)P_l(x) dx\text{.}
\end{equation}
This is convenient because it implies that the angular portion of Eq. \ref{eq:C3Basic} can represent any function through a series of Legendre polynomials.

The only task remaining is choosing $\beta$, $\hat{C}_2$, $R$, and $\alpha_l$ in order to produce the targeted structure. The $\beta$ constant is not strictly necessary since changing its value from unity is equivalent to modifying $R$. For convenience, however, $\beta$ was introduced in order to easily tune the relative strengths of the two- and three-point interactions. Below we motivate the parameters choices listed in the Supplementary Materials by considering diamond cubic and disordered CaF$_2$ as examples.

We first consider the diamond cubic crystal structure. Since diamond cubic, like all crystal structures, is periodic, the density field can be expanded in a Fourier series, i.e.,
\begin{equation} \label{eq:exp}
n(\vect{r}) = \bar{n} + \sum_j A_j e^{i \vect{k}_j \cdot \vect{r}} \text{,}
\end{equation}
where $\bar{n}$, the average value of $n$, will be set to zero in all cases in this Letter for simplicity. When the diamond cubic structure is expressed on a simple cubic lattice, the atoms are located at both the fcc sites and the fcc sites translated by $(1/4, 1/4, 1/4)$, for a total of eight atoms per unit cell. If $\vect{k}$ is then expressed in terms of primitive reciprocal lattice vectors, i.e. $\vect{k} = \mh\uv{b}_1 + \mk\uv{b}_2 + \ml\uv{b}_3$, the amplitudes for an atomic density represented by Dirac delta functions at the atomic positions are
\begin{eqnarray} \label{eq:strFactor}
  A_{j(\mh \mk \ml)} =
  \begin{cases}
    8 & \text{if } \mh + \mk + \ml = 4N \\
      & \text{and $\mh$, $\mk$, $\ml$ are all even} \\
    4(1 + i) & \text{if } \mh + \mk + \ml = 4N + 1 \\
              & \text{and $h$, $k$, $l$ are all odd} \\
    4(1 - i) & \text{if } \mh + \mk + \ml = 4N + 3 \\
    			 & \text{and $\mh$, $\mk$, $\ml$ are all odd} \\
    0 & \text{otherwise}
  \end{cases}
\end{eqnarray} 
where $N$ is an integer. 

The calculated amplitudes (Eq. \ref{eq:strFactor}) are used to select the parameters of the model. First, we discuss the two-point correlation. Since the smallest set of reciprocal lattice vectors with nonzero amplitudes in Eq. \ref{eq:strFactor} is the \{111\} set, we let
\begin{equation} \label{eq:xpfcC2}
\hat{C}_2(k) \equiv A_2 e^{-\frac{(k-q_1)^2}{2\sigma^2}} \text{,}
\end{equation} 
where $q_1 = 2 \pi \sqrt{3}/a_0$, $A_2$ is a temperature-dependent parameter, $a_0$ is the lattice constant, and $\sigma$ is related to interfacial free energy, consistent with structural PFC  (XPFC) models \cite{Greenwood2010, Greenwood2011, Chan2015, Alster2017}. 

To choose the parameters for the three-point correlation, consider the free energy  resulting from it
\begin{equation} \label{eq:fe3}
F_3/V = -\frac{1}{6} \sum_{pqr} \hat{C}_3(k_p,k_q,\uv{k}_p \cdot \uv{k}_q) A_p A_q A_r \delta_{\vect{k}_p + \vect{k}_q + \vect{k}_r, \vect{0}} \text{.}
\end{equation}
Notice that the only nonzero contributions to this energy come from groups of vectors, $[\vect{k}_p, \vect{k}_q, \vect{k}_r]$, that satisfy both $\hat{C}_3(k_p,k_q,\uv{k}_p \cdot \uv{k}_q) \neq 0$ and $\vect{k}_p + \vect{k}_q + \vect{k}_r = \vect{0}$. From Eq. \ref{eq:C3Basic}, it is clear that $\hat{C}_3$ is nonzero only when both $R(k_p)$ and $R(k_q)$ are nonzero. Consequently, $R(k)$ can be interpreted as a weighting factor for wave vector magnitudes, like $\hat{C}_2$. Therefore, it is convenient to define $R$ in a similar manner as we define $\hat{C}_2$. For diamond and most other crystal structures, we found that $R(k) = \hat{C}_2(k)$ (given by Eq. \ref{eq:xpfcC2}) works well. In the limit of small $\sigma$ for this choice of $R$, only groups where the first two wave vectors are of magnitude $q_1$ can contribute to the three-point term of the free energy. Since the $\vect{k}_p + \vect{k}_q + \vect{k}_r = \vect{0}$ condition must also be satisfied, only groups like
$[(111), (111), (\bar{2}\bar{2}\bar{2})]$, 
$[(111), (11\bar{1}), (\bar{2}\bar{2}0)]$, and
$[(111), (1\bar{1}\bar{1}), (\bar{2}00)]$ contribute to the free energy.
However, wave vectors of type $(\bar{2}\bar{2}\bar{2})$ and $(\bar{2}00)$ have zero amplitude for the diamond structure (see Eq. \ref{eq:strFactor}). This leaves only groups equivalent to $[(111), (11\bar{1}), (\bar{2}\bar{2}0)]$ as contributors to the three-point term (e.g., $[(1\bar{1}\bar{1}), (\bar{1}\bar{1}\bar{1}), (022)]$ would be another example of a contributing group). It can be shown that for these groups, $\uv{k}_p \cdot \uv{k}_q = 1/3$, or equivalently, the angle between the $p$ and $q$ planes is $\cos^{-1}(1/3) \simeq 70.5 \degsym$. Also note that, for these groups, the product $A_p A_q A_r$ is always positive (see Eq. \ref{eq:strFactor}). Consequently, by choosing the coefficients $\alpha_l$ in Eq. \ref{eq:C3Basic} in such a way that $\hat{C}_3$ is positive when $\uv{k}_p \cdot \uv{k}_q = 1/3$ and zero otherwise, we energetically promote the angle $\cos^{-1}(1/3)$, corresponding to the angle between $\{111\}$ planes. One simple way to do so is for $B$ in Eq. \ref{eq:alphalGen} to be a delta function centered at $x = 1/3$. Namely, we take
\begin{equation} \label{eq:alphalDiamond}
\alpha_l = \frac{2l+1}{2} \int_{-1}^{1} \delta(x - 1/3)P_l(x) dx = \frac{2l+1}{2}P_l(1/3)\text{.}
\end{equation}
Trial-and-error is required for determining how many terms are necessary. For this case, we found that $l_{\text{max}} = 3$ was sufficient. To demonstrate that diamond cubic is likely the equilibrium structure, it was tested against bcc, fcc, simple cubic, hexagonal rods, hcp, disordered CaF$_2$, graphene rods, simple hexagonal, simple cubic rods, and stripes \cite{Chan2015}. To test for the equilibrium phase, an initial condition is set up so that it approximates a possible structure in a unit cell of the appropriate size, and then the energy is minimized through standard conserved nonlocal dynamics \cite{Mellenthin2008},
\begin{equation} \label{eq:evol}
\frac{\partial n}{\partial t} = -\frac{\delta F}{\delta n} + \frac{1}{V} \int_V \frac{\delta F}{\delta n} d\vect{r} \text{.}
\end{equation}
Out of the structures that were tested, diamond cubic was the one with the lowest energy. However, because only a finite number of structures can be examined, this does not prove that the global minimum energy structure was found. Nonetheless, it was also observed that if a system of size $4 \times 4 \times 4$ unit cells is initialized with noise, a diamond cubic structure forms. Although using purely the dynamics of Eq. \ref{eq:evol} results in the structure becoming kinetically trapped in a high-energy, low-amplitude state (i.e., the evolution toward equilibrium is very slow), the dynamics can be accelerated by multiplying the amplitude of the high-energy structure by a large factor (on order of $500$), after which the diamond cubic phase quickly appears when the system is relaxed, regardless of the seed used to generate the initial random condition.  The formation of the diamond structure without any \textit{a priori} information about the equilibrium state, except through the periodic boundary conditions, suggests that there are no unaccounted for lower energy phases.

\begin{figure}
    \includegraphics[width=\linewidth]{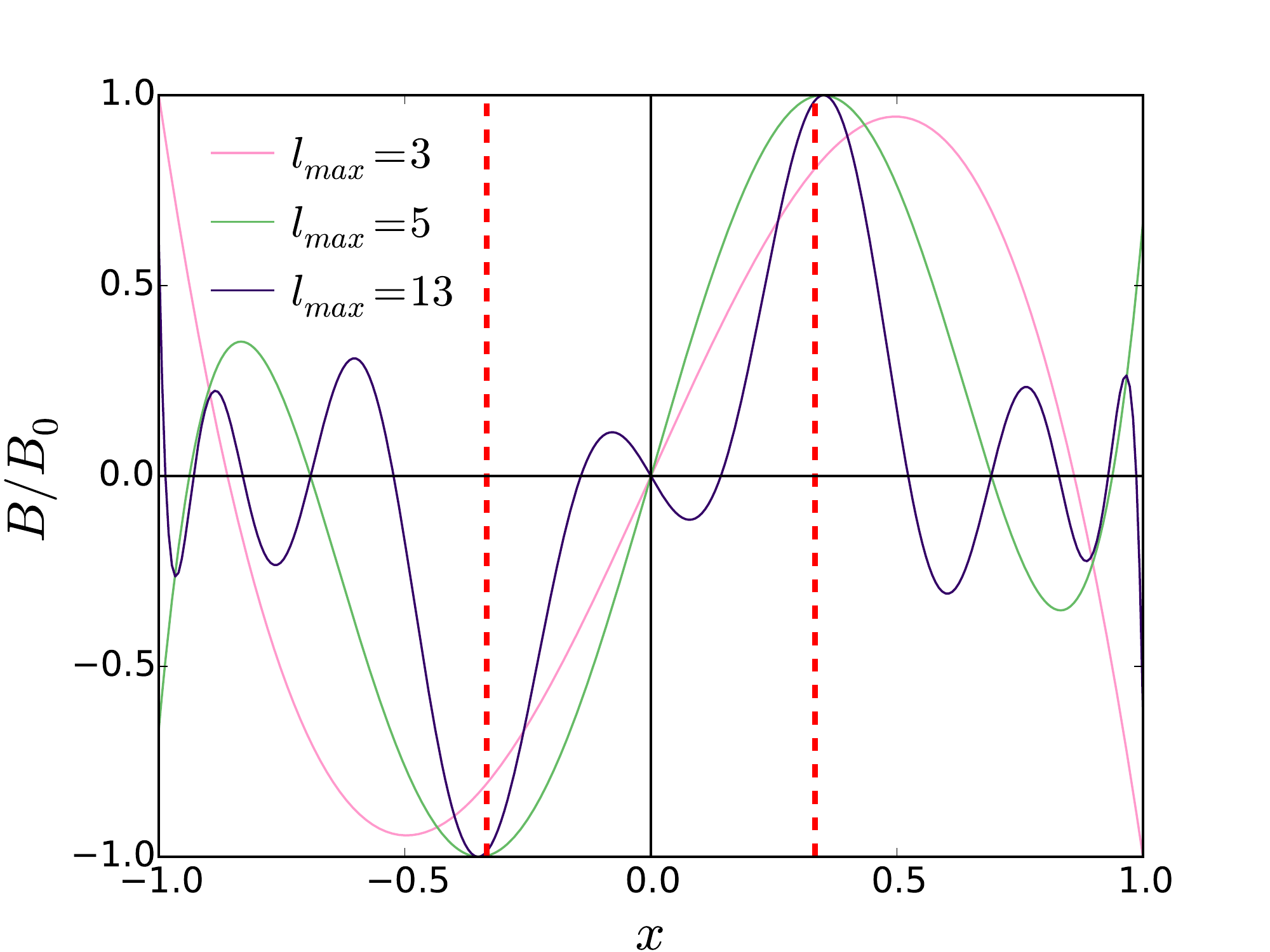}
    \caption{Plot of $B(x)$ for $l_{\text{max}} = 3$, $5$, and $13$, normalized to their maximum values. For $l_{\text{max}} < 5$, the functions' peaks are not close to $\pm 1/3$, marked by the vertical dashed lines. Although the peaks for $l_{\text{max}} = 5$ are on target, the wide breadth of peaks includes the values $\pm 1/2$. In contrast, $l_{\text{max}} = 13$ has sharp, centered peak with close to zero baseline.}
    \label{fig:BPlot}
\end{figure}

As a second example, we present the case of a single-component CaF$_2$ model. Consider a simple cubic lattice with atoms at the fcc positions and at the tetrahedral voids. The amplitudes for this structure are
\begin{eqnarray} \label{eq:strFactorCaF2}
  A_{j(\mh \mk \ml)} =
  \begin{cases}
    12 & \text{if } \mh + \mk + \ml = 4N \\
      & \text{and $\mh$, $\mk$, $\ml$ are all even} \\
    4 & \text{if $\mh$, $\mk$, $\ml$ are all odd} \\
    -4 & \text{if } \mh + \mk + \ml = 4N + 2 \\
    	   & \text{and $\mh$, $\mk$, $\ml$ are all even} \\
    0 & \text{otherwise.}
  \end{cases}
\end{eqnarray} 
Notice that this structure has nonzero amplitudes for the same $(\mh \mk \ml)$ (i.e., has the same extinction symbol \cite{Aroyo2016}) as fcc but with different amplitude values. Since the energy from the two-point interaction term is a function of only the magnitudes of the amplitudes and not their phase, it is difficult, and maybe impossible, to generate a two-point correlation that is able to stabilize this structure over fcc and bcc. However, discerning between the fcc and CaF$_2$ structures is possible using a three-point correlation. Like for diamond cubic, $R(k) = A_2 e^{-\frac{(k-q_1)^2}{2\sigma^2}}$ where $q_1 = 2 \pi \sqrt{3}/a_0$. This selects the first nonzero reciprocal lattice vector (the $\{111\}$ planes). Using this $R$, there are groups with two relevant angles, unlike the diamond cubic case: $[(111), (11\bar{1}), (\bar{2}\bar{2}0)]$ type groups with $\vect{k}_p \cdot \vect{k}_q = 1/3$ and $[(111), (1\bar{1}\bar{1}), (\bar{2}00)]$ groups with $\vect{k}_p \cdot \vect{k}_q = -1/3$. For the former, $A_pA_qA_r > 0$, and for the latter, $A_pA_qA_r < 0$. Consequently, Eq. (\ref{eq:alphalGen}) becomes 
\begin{equation}
\alpha_l = \frac{2l + 1}{2}(-P_l(-1/3) + P_l(1/3)) \text{.}
\end{equation}
For this case, $l_{\text{max}} = 5$ was sufficient for $B(x)$ to produce a peak at $\pm 1/3$ (see Fig. \ref{fig:BPlot}). However, among the structures examined, the lowest energy state with this $l_{\text{max}}$ was found to be an ``inverse'' bcc structure (i.e., $-n$ has a bcc structure), rather than the single-component CaF$_2$ structure. This occurs because inverse bcc has contributing groups equivalent to $[(110), (\bar{1}0\bar{1}), (0\bar{1}1)]$, like bcc, and these groups have $\vect{k}_p \cdot \vect{k}_q = -1/2$. When $l_{\text{max}} = 5$, $B(x)$ has a broad peak and $B(-1/2) \simeq B(-1/3)$, so the same symmetry reasons that normally prefer bcc over fcc in this case prefer inverse bcc over the single-component CaF$_2$ structure \cite{Alexander1978}. Consequently, the peaks were narrowed with $l_{\text{max}} = 13$, in which case the CaF$_2$ structure is the energy minimum among all structures examined. Although $l_{\text{max}} = 13$ at first might appear computationally expensive, it can be evaluated efficiently because every convolution term can be computed in parallel.

A similar approach was employed to identify the parameters for the simple hexagonal, simple cubic, and $X_3$ structures, in addition to graphene layers. All crystal structures were found to be lower in energy than all the compounds listed previously in connection with diamond cubic. Additionally, CaF$_2$ spontaneously ordered from noise, and the rest (except $X_3$, which spontaneously ordered to a higher energy phase) ordered from noise with the aforementioned method to accelerate the kinetics. A full listing of parameters used is contained in the Supplementary Materials.

\begin{figure}  
    \includegraphics[width=\linewidth]{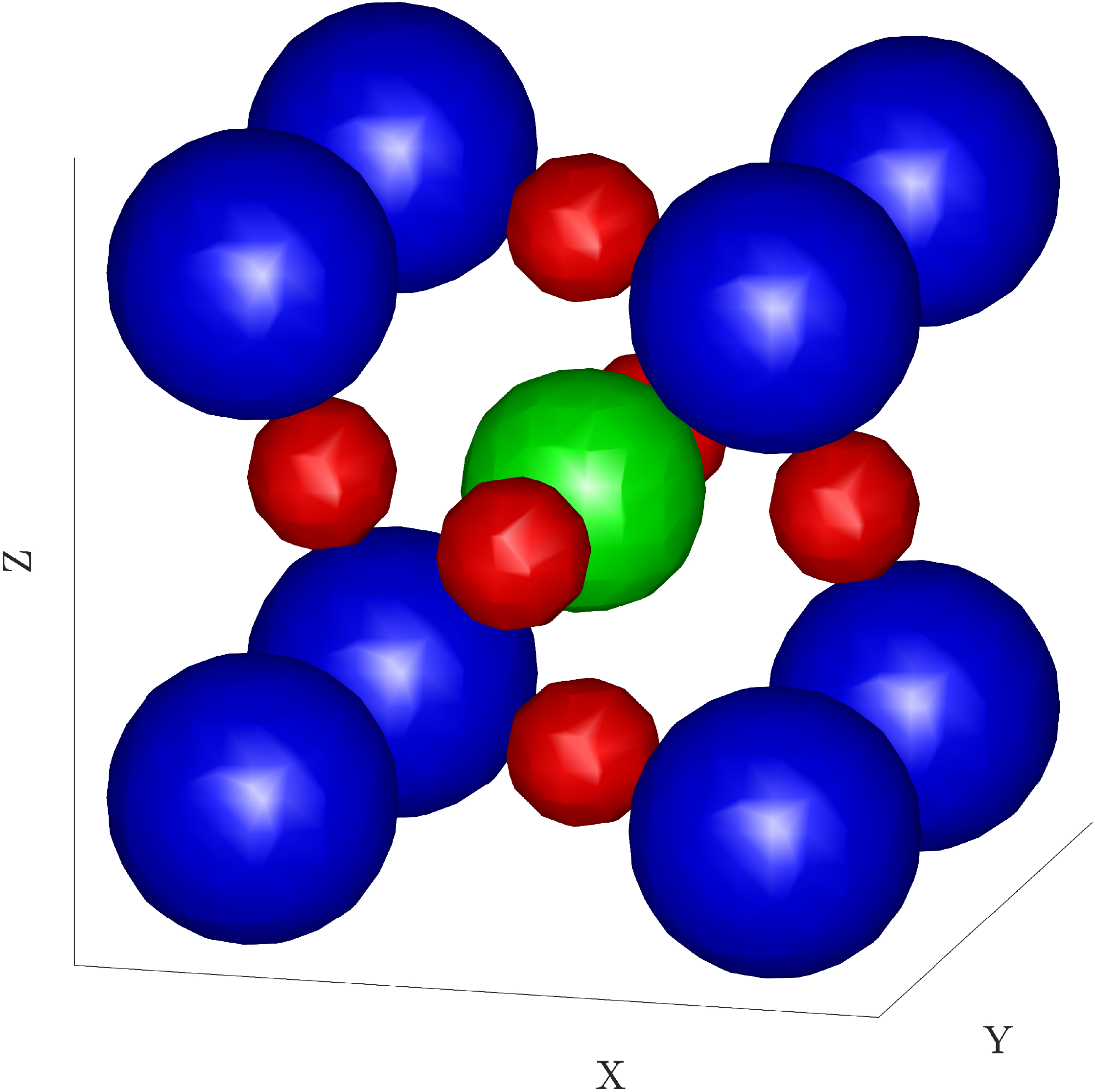}
    \caption{Three-dimensional isosurface plot of equilibrated cubic perovskite using Eq. \ref{eq:ps}. For the canonical $ABX_3$ perovskite, blue $A$ atoms are at the corners and a green $B$ atom is at the center surrounded by six red $X$ nearest-neighbors.}
    \label{fig:ps1}
\end{figure}

As a capstone demonstration, we show how the single-component three-point correlations can be combined to construct a simple PFC model for perovskite (Fig. \ref{fig:ps1}) where the only interaction coupling the components is an excluded volume term. Since the model does not include electrostatic interactions, the structure is equivalent to antiperovskite as well. The free energy of this model is given by
\begin{align} \label{eq:ps}
F[n_A, n_B, n_X] = F_A[n_A] + F_B[n_B] + F_X[n_X] \nonumber \\
+ Z \int_V (n_A n_B + n_A n_X + n_B n_X) d\vect{r} \text{,}
\end{align}
where $F_A$ and $F_B$ are simple cubic single-component free energies, $F_X$ is the free energy for $X_3$, and $Z > 0$. Note that each of the single-component free energies are of the form given by Eq. \ref{eq:esum}. If the parameters for $A$ and $B$ are the same, there is no driving force for the $B$ atoms, rather than the $A$ atoms, to have a coordination number of six. To break this symmetry, the radii of the $B$ isosurfaces were made smaller (as in the actual perovskite structure). A small $Z$ was found to be sufficient for the perovskite structure to be an energy minimum and be able to spontaneously order from noise (parameters are given in Supplementary Materials).

There are many potential ways in which the method introduced above can be applied in future research. For all of the particular structures described, information including their elastic properties, surface energies, and grain boundary morphologies are of interest. For example, the perovskite model could be used to model chemical vapor deposition grown perovskite solar cells \cite{Tavakoli2015}. Other potential applications include combining our model with the PFC ordering model \cite{Alster2017} to create a two-component CaF$_2$ model for modeling $\theta'$ precipitates in Al-Cu alloys, combining it with a vapor phase model \cite{Schwalbach2013, Kocher2015} to create a 3D single layer graphene model, and extending it to other complex phases such as Heusler alloys and Laves phases.

\begin{acknowledgments}
Boaz Haberman, Ken Elder, Zhifeng Huang, Matthew Seymour, and Jason Luce are thanked for many helpful discussions. The authors are also grateful to Thomas Cool for visualization assistance. The National Science Foundation Graduate Research Fellowship, NSF DMR-1507033, NSF DMR-3003700315, and Northwestern University are thanked for financial support. 
\end{acknowledgments}

\bibliography{Point3a_paper}

\end{document}


\title{Crystal Structure Parameters}

\begin{table*}
  \caption{Parameter values that were used to stabilize various crystal structures where 
$
\hat{C}_2(k) = A_0 e^{-\frac{k^2}{2 \sigma^2}} + \max_i(A_i e^{-\frac{(k - q_i)^2}{2 \sigma^2}})
$
and 
$
\hat{R}(k) = \max_j(A_j e^{-\frac{(k - r_j)^2}{2 \sigma^2}})
$. The compound $X_3$ refers to the structure of the $X$ component of the $ABX_3$ perovskite/anti-perovskite structure. The asterisk denotes that for graphene layers, $\alpha_0$ was changed from $-1/2$, inconsistent with the $P_l$ calculation. This change was motivated by the fact that the graphene layers structure is the inverse (negative peaks) of simple hexagonal.\label{table:main}
}
  \centering
\begin{ruledtabular}
\begin{tabular}{l | l l l l l l}
 & \textbf{Diamond Cubic} & \textbf{Simple Hexagonal} & \textbf{Simple Cubic} & \textbf{Graphene Layers} & $\mathbf{CaF_2}$ & $\mathbf{X_3}$\\
\hline
$A_0$ & - & - & - & - & $-5$ & -\\
$A_1$ & $1.0$ & $1.0$ & $1.0$ & $1.0$ & $0.9$ & $0.95$\\
$A_2$ & - & $0.99$ & $0.98$ & $0.99$ & $1.05$ & $0.95$\\
$A_3$ & - & $0.98$ & - & $0.98$ & - & $1.0$\\
$A_4$ & - & - & - & - & - & $0.98$\\
$q_1$ & $2\pi$ & $2\pi$ & $2\pi$ & $2\pi $ & $2\pi \sqrt{3/8}$ & $2\pi$\\
$q_2$ & - & $3\pi/\sqrt{8}$ & $2\pi \sqrt{2}$ & $3\pi/\sqrt{8}$ & $2\pi$ & $2\pi \sqrt{2}$\\
$q_3$ & - & $2\pi \sqrt{41/32}$ & - & $2\pi \sqrt{41/32}$ & - & $2\pi \sqrt{3}$\\
$q_4$ & - & - & - & - & - & $4\pi$\\
$r_1$ & $2\pi$ & $2\pi$ & $2\pi$ & $2\pi$ & $2\pi \sqrt{3/8}$ & $2\pi$\\
$r_2$ & - & $3\pi/\sqrt{8}$ & $2\pi \sqrt{2}$ & $3\pi/\sqrt{8}$ & - & -\\
$r_3$ & - & $2\pi \sqrt{41/32}$ & - & $2\pi \sqrt{41/32}$ & - & -\\
$\beta$ & $1.3$ & $1.0$ & $1.0$ & $1.0$ & $0.8$ & $1.3$\\
$\sigma$ & $0.1$ & $0.1$ & $0.1$ & $0.1$ & $0.1$ & $0.1$\\
$\frac{2 \alpha_l}{2 l + 1}$ & $P_l(1/3)$ & $P_l(-13/17)$ & $P_l(0)$ & $-P_l(-13/17)$ & $P_l(1/3) - P_l(-1/3)$ & $-P_l(0)$ \\
$\alpha_0$ & $1/2$ & $1/2$ & $1/2$ & $-1/6^*$ & - & $-1/2$\\
$\alpha_1$ & $1/2$ & $-39/34$ & - & $39/34$ & $1$ & -\\
$\alpha_2$ & $-5/6$ & $545/578$ & $-5/4$ & $-545/578$ & - & $5/4$\\
$\alpha_3$ & $-77/54$ & $1001/9826$ & - & $-1001/9826$ & $-77/27$ & -\\
$\alpha_4$ & - & $-241911/167042$ & $27/16$ & $241911/167042$ & - & $-27/16$\\
$\alpha_5$ & - & - & - & - & $11/3$ & -\\
$\alpha_6$ & - & - & - & - & - & -\\
$\alpha_7$ & - & - & - & - & $-605/243$ & -\\
$\alpha_8$ & - & - & - & - & - & -\\
$\alpha_9$ & - & - & - & - & $-9101/19683$ & -\\
$\alpha_{10}$ & - & - & - & - & - & -\\
$\alpha_{11}$ & - & - & - & - & $229057/59049$ & -\\
$\alpha_{12}$ & - & - & - & - & - & -\\
$\alpha_{13}$ & - & - & - & - & $-353807/59049$ & -\\
$a_1$ & $\sqrt{3}$ & $2/\sqrt{3}$ & $1$ & $2/\sqrt{3}$ & $\sqrt{8}$ & $1$\\
$a_2$ & $\sqrt{3}$ & $2$ & $1$ & $2$ & $\sqrt{8}$ & $1$\\
$a_3$ & $\sqrt{3}$ & $4 \sqrt{2}/3$ & $1$ & $4 \sqrt{2}/3$ & $\sqrt{8}$ & $1$
\end{tabular}
\end{ruledtabular}
\end{table*}

For perovskite, nearly all parameters are the same as for the single components listed in Table \ref{table:main}. The only changes are that for $B$, $A_1 = 0.95$, $A_2 = 0.93$, and $\beta = 0.95$. Additionally, $Z = 0.05$.



\section*{Proof that $\mathbf{C^{(lm)}}$ is real}
To prove $C^{(lm)}(r, \uv{r})$ is real, use a plane wave expansion
\begin{equation}
e^{i\vect{k} \cdot \vect{r}} = 4\pi \sum_{l=0}^{\infty} \sum_{m=-l}^{l} i^l j_l(k r)Y_{lm}(\uv{k})Y_{lm}(\uv{r})
\end{equation}
where $j_l$ are the spherical Bessel functions. Consequently,
\begin{align}
C^{(lm)}(r, \uv{r}) =& \left ( \frac{1}{2\pi} \right )^3 (-i)^l \sqrt{\frac{4 \pi}{2l+1}} \beta
\int R(k)Y_{lm}(\uv{k}) e^{i\vect{k} \cdot \vect{r}} d\vect{k} \\
=& Y_{lm}(\uv{r}) \left ( \frac{1}{2\pi} \right )^3 \sqrt{\frac{4 \pi}{2l+1}}\beta 4\pi \int_0^{\infty} R(k) j_l(k r) k^2 dk \label{eq:ClmReal}
\end{align}
by orthogonality of spherical harmonics. Eq. \ref{eq:ClmReal} is real since $Y_{lm}$, $R$, and $j_l$ are real.